\documentclass[prl,twocolumn,showpacs,amsmath,amssymb]{revtex4}

\usepackage{graphicx}
\usepackage{dcolumn}
\usepackage{bm}

\begin{document}

\preprint{APS/123-QED}

\title{Fermi liquid state and enhanced electron correlations\\ in the new iron pnictide CaFe$_4$As$_3$}

\author{Liang L. Zhao,$^{1}$ Tanghong Yi,$^{2}$ James C. Fettinger,$^2$ Susan M. Kauzlarich$^{2}$ and E. Morosan$^{1}$}
\affiliation{$^1$Department of Physics and Astronomy, Rice University, Houston TX 77005 \\ $^2$Department of Chemistry, One Shields Ave, U.C. Davis, Davis, CA 95616}

\date{\today}

\begin{abstract}
The newly discovered CaFe$_4$As$_3$ system displays low-temperature Fermi liquid behavior, with enhanced electron-electron correlations. At high temperatures, the magnetic susceptibility shows Curie-Weiss behavior, with a large temperature-independent contribution. Antiferromagnetic ordering is observed below T$_N$ = (88.0 $\pm$ 1.0) K, possibly via a spin density wave (SDW) transition. A remarkably sharp drop in resistivity occurs below  T$_2$ = (26.4 $\pm$ 1.0) K, correlated with a similarly abrupt increase in the susceptibility, but no visible feature in the specific heat. The electronic specific heat coefficient $\gamma$ at low temperatures is close to 0.02 J mol$^{-1}_{Fe}$ K$^{-2}$, but a higher value for $\gamma$ ($\sim$0.08 J mol$^{-1}_{Fe}$ K$^{-2}$ can be inferred from a linear C$/$T \textit{vs.} T$^2$ just above T$_2$. The Kadowaki-Woods ratio A$/\gamma^2$ = 55$*$10$^{-5}$ $\mu~\Omega$cm mol$^2$ K$^2 $mJ$^{-2}$ is nearly two orders of magnitude larger than that of heavy fermions.
\end{abstract}


\maketitle

\begin{figure}[b]
\begin{center}
\includegraphics[width=2.8in]{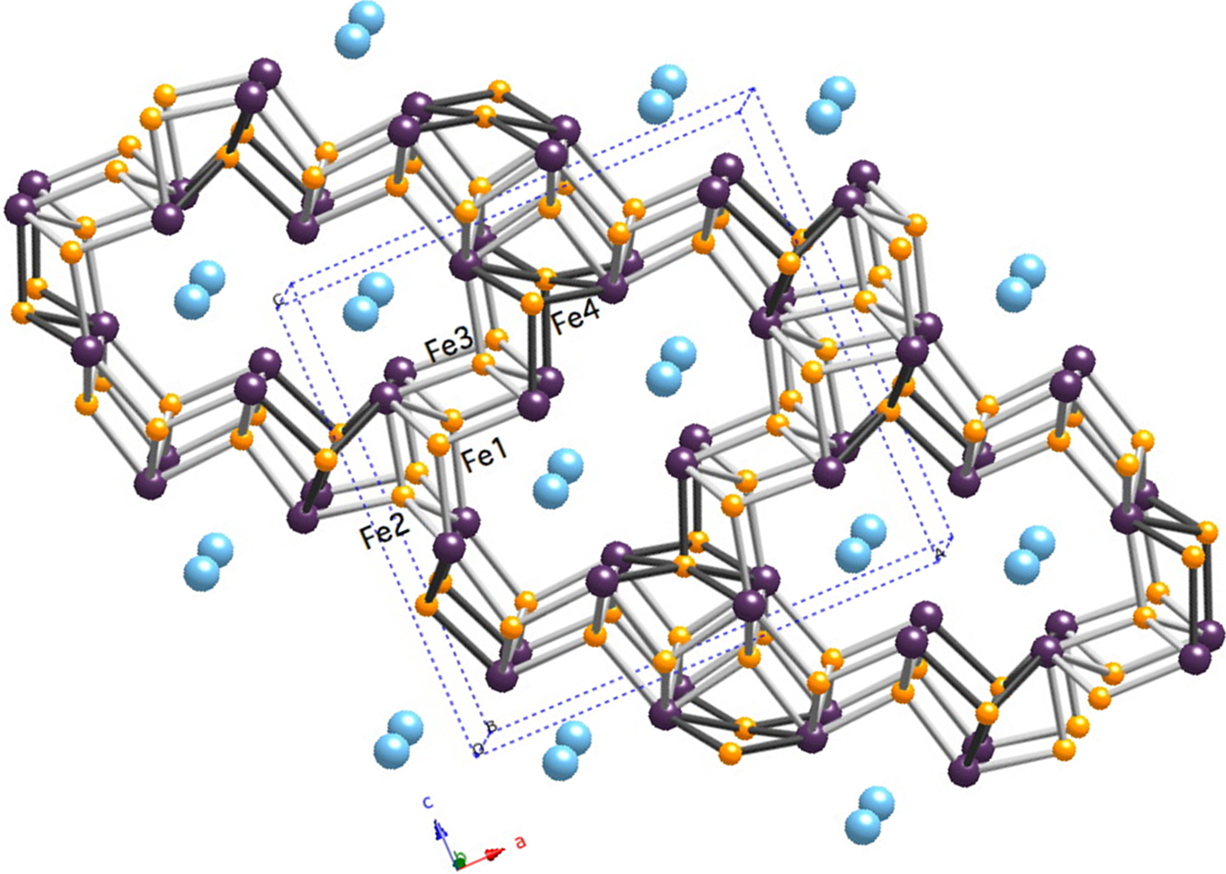}
\caption{ac-plane view of the CaFe$_4$As$_3$ structure, with the Ca, Fe and As represented by blue (large), orange (small) and purple (medium) spheres, respectively. 4-fold coordinated Fe and As atoms form buckled sheets (grey bonds), similar to Fe-As planes in Fe-pnictide superconductors; these fold back into each other around the 5-fold coordinated Fe4 atoms.} \label{crystal01}
\end{center}
\end{figure}

Many advances in science in general, and in condensed matter in particular, have been brought about by the discovery of new materials. The most recent example are the high temperature iron-oxypnictide superconductors \cite{kam08jacs}, a case of accidental incorporation of oxygen in a rare earth-iron-pnictide ternary via tin flux. After the original discovery of superconductivity in La(O$/$F)FeAs, the past year has seen a revival in the interest in superconductivity, fueled by the discovery of additional classes of homologous superconductors based on BaFe$_2$As$_2$ \cite{Ba122,BaKFeAs122}, FeSe \cite{FeSe} and LiFeAs \cite{LiFeAs2,LiFeAs4}, and an increase in T$_c$ to new records (over 50 K) for non-cuprate materials. Here we report the discovery of a new Fe-As compound, CaFe$_4$As$_3$, with complex electronic and magnetic properties despite lack of superconductivity down to 1.8 K. CaFe$_4$As$_3$ is a new framework structure composed of shared Fe-As tetrahedra, with Ca atoms (blue/large spheres, Fig. \ref{crystal01}) sitting in channels defined by these shared tetrahedra. Two phase transitions are observed in this material, an antiferromagnetic ordering below T$_N~\approx$ 88.0 K, likely associated with a spin density wave (SDW) transition, and a lower transition around T$_2~\approx$ 26.4 K. Fermi liquid behavior with enhanced electron correlations is observed below this lower transition. As a result, the Kadowaki-Woods (KW) ratio is nearly two orders of magnitude larger than the corresponding value for heavy fermions, with a correspondingly enhanced Wilson ratio (R).

Single crystals of CaFe$_4$As$_3$ have been synthesized using tin flux, starting from a composition Ca:Fe:As:Sn = 1:4:3:40. The constituent elements were placed in evacuated silica tubes, slowly heated up to 1100 $^0$C, followed by cooling at a rate of 4.5  $^0$C/hour down to 600 $^0$C. After the tin flux was decanted, well-formed, thin rods were obtained, typically about 1 mm$^2$ in cross-section, and up to 8 mm long. Anisotropic magnetization measurements were performed using the Reciprocating Sample Option (RSO) of the Quantum Design Magnetic Property Measurement System (QD MPMS). Specific heat measurements were performed in a Quantum Design Physical Property Measurement System (QD PPMS), using an adiabatic relaxation technique. Given the rod-like geometry of the sample, the AC resistivity measurements (\textit{i} = 1 mA, f = 17.77 Hz) were limited to those for current along the rod axis, and they were done using a standard four probe method in the same QD PPMS environment. The structure of CaFe$_4$As$_3$ was determined by single-crystal X-ray diffraction. Needle shape crystals were cut to size ($\sim$0.22 $\times$ 0.11 $\times$ 0.08 mm$^3$) and placed in a nitrogen stream. Diffraction data were acquired using a Bruker SMART 1000 CCD diffractometer utilizing a graphite-monochromatic Mo K$\alpha$ radiation ($\lambda$ = 0.71073 ${\AA}$) at T = 90(2) K and the SMART software package. SAINT was employed for data frame integrations. SADABS was used to correct the data for Lorentz and polarization effects and to apply a numerical absorption correction. XPREP was used to identify the space group and to create the data files. The structure was determined by direct methods with SHELXS and refined with SHELXL \cite{xray}.

Above 90 K, CaFe$_4$As$_3$ is orthorhombic (space group Pnma) with lattice parameters a = 11.8840(7) ${\AA}$, b = 3.7342(2) ${\AA}$, and c = 11.5857(7) ${\AA}$. There are four crystallographic sites for Fe (small spheres, Fig. \ref{crystal01}) and several close Fe-Fe distances, between 2.5952(6) ${\AA}$ and 2.8629(4) ${\AA}$. The Fe1-Fe3 atoms are four-fold coordinated with As (medium spheres) (grey lines, Fig. \ref{crystal01}), giving rise to buckled Fe-As sheets similar to the Fe-pnictide planes in the new superconductors. These sheets fold back into each other around the five-fold coordinated Fe4 atoms, forming channels along the b axis, with the Ca atoms (large spheres, Fig. \ref{crystal01}) sitting in these channels.

\begin{figure}[t]
\begin{center}
\includegraphics[width=2.6in]{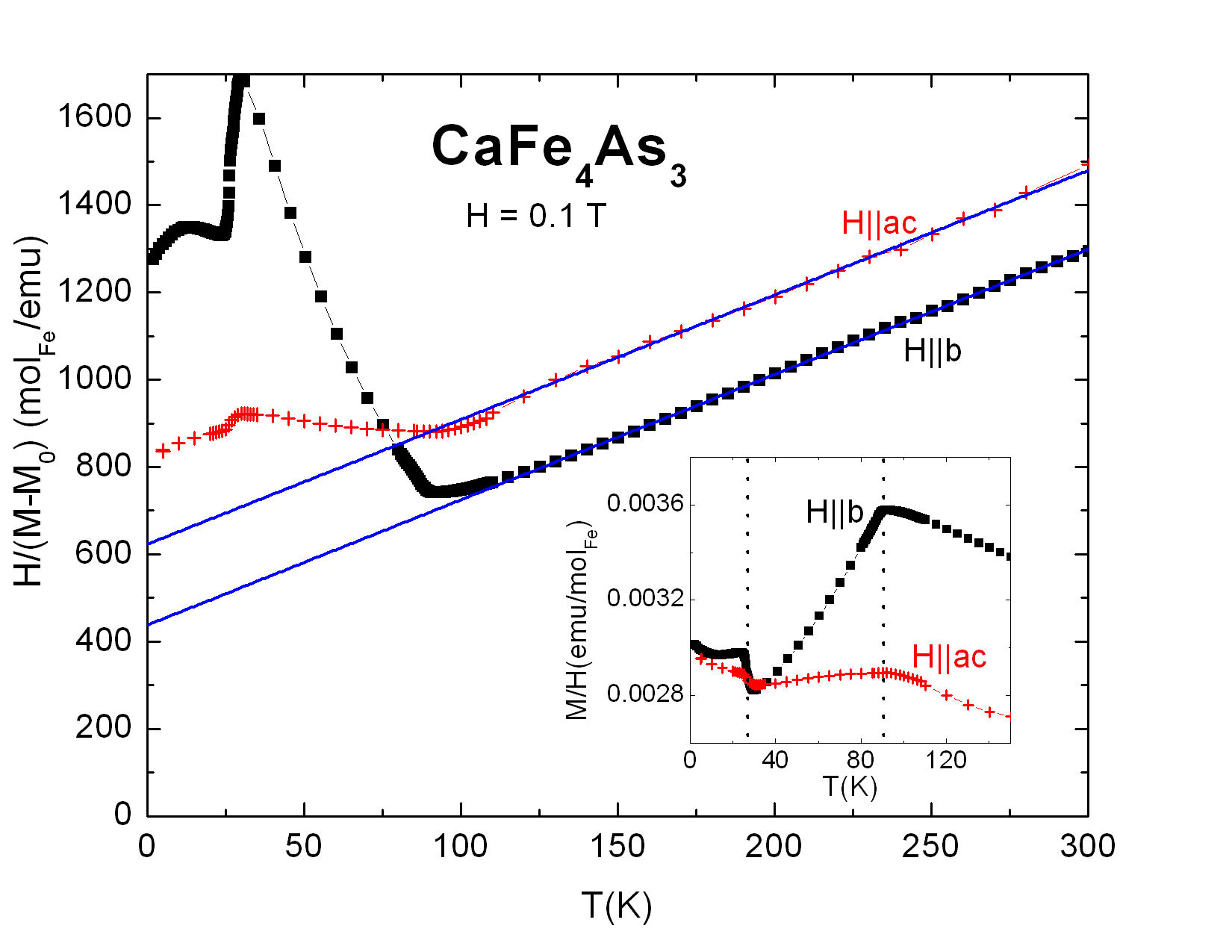}
\caption{Anisotropic inverse magnetic susceptibility (H = 0.1 T) for H$\parallel$ac (crosses) and H$\parallel$b (full squares). The straight lines represent the high temperature fits to the Curie-Weiss law, after a constant $\chi_0$ has been taken into account. Inset: low temperature anisotropic $\chi(T)$ showing a sharp decrease below T$_N$ = 87.0 K, and an upturn around the low-T transition T$_2$ = 26.4 K.} \label{MT}
\end{center}
\end{figure}

The anisotropic inverse susceptibility $1/\chi(T)=H/M(T)$ is shown in Fig. \ref{MT}, for an applied magnetic field H $=$ 0.1 T. At high temperatures ($T~>~130~K$) the susceptibility (Fig. \ref{MT}) can be fit to a Curie-Weiss law $\chi(T)~=~\chi_0~+~C~/~(T~-~\theta)$, after a temperature-independent contribution $\chi_0$ has been accounted for. Larger $\chi$(T) values are observed for $H~||~b$ (black squares, inset Fig. \ref{MT}) than for H$||$ac (red crosses, inset Fig. \ref{MT}). The $\chi_0$ values sum up the core diamagnetic, Pauli and Landau contributions, and are determined from the data fits to be 2.23$*$10$^{-3}$ emu mol$^{-1}_{Fe}$ and 1.76$*$10$^{-3}$ emu mol$^{-1}_{Fe}$, for H$||$b and H$||$ac respectively. Such large $\chi_0$ values strongly suggest enhanced Pauli paramagnetism (since the core diamagnetism is typically of the order of 10$^{-6}~-~$ 10$^{-5}$ emu mol$^{-1}$ \cite{carlin}, and $\chi_{Landau}\approx-1/3~\chi_{Pauli}$). An effective moment $\mu_{eff}~=~$1.66 $\mu_B/$Fe is determined for both field orientations, much smaller than the theoretical values expected for either Fe$^{2+}$ (4.9 - 6.7 $\mu_B/$Fe) or Fe$^{3+}$ (5.9 $\mu_B/$Fe) \cite{ashcroft}. Although CaFe$_4$As$_3$ is not superconducting above 1.8 K, nor does it have the layered Fe - As structure ubiquitous in the new Fe superconductors, the structure is comprised of Fe-As buckled sheets, similar to the infinite planes in the latter compounds. The reduced moment per Fe ion is also reminiscent of the Fe superconductors: 0.4 - 0.8 $\mu_B/$Fe has been reported in the ROFeAs compounds \cite{La1111a}, and close to 0.87 $\mu_B/$Fe for BaFe$_2$As$_2$ \cite{huang122}. One more notable similarity with the superconducting Fe-pnictides is an antiferromagnetic-like transition which occurs in CaFe$_4$As$_3$ around T$_N~=$ 88.0 K, very close to the spin density wave (SDW) transition temperature observed in single crystals of BaFe$_2$As$_2$ \cite{nini01}.

\begin{figure}[t]
\begin{center}
\includegraphics[width=2.6in]{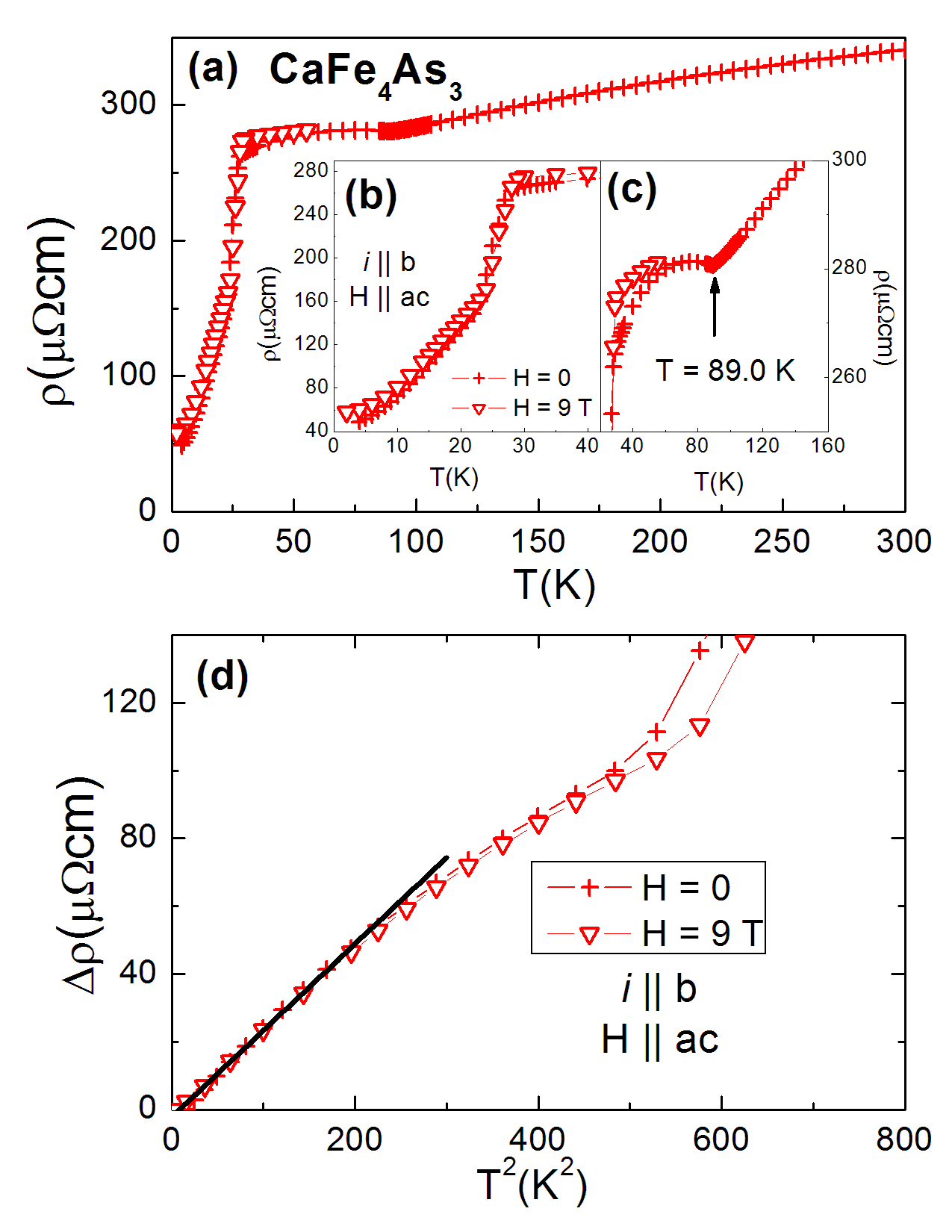}
\caption{Resistivity of CaFe$_4$As$_3$ for \textit{i}$\parallel$b, and H = 0 (crosses) and H = 9 T (triangles) (H$\perp$\textit{i}). Details around the (b) T$_2$ and (c) T$_N$ transitions. (d) Low temperature $\Delta\rho=\rho-\rho_0$ \textit{vs.} T$^2$, with the residual resistivity $\rho_0~\approx$ 42 $\mu\Omega$cm as determined from (b); straight line is a fit to $\Delta\rho=\rho-\rho_0$ = AT$^2$, which gives A = 0.25 $\mu\Omega$ cm K$^{-2}$} \label{RT}
\end{center}
\end{figure}

\begin{figure}[t]
\begin{center}
\includegraphics[width=2.8in]{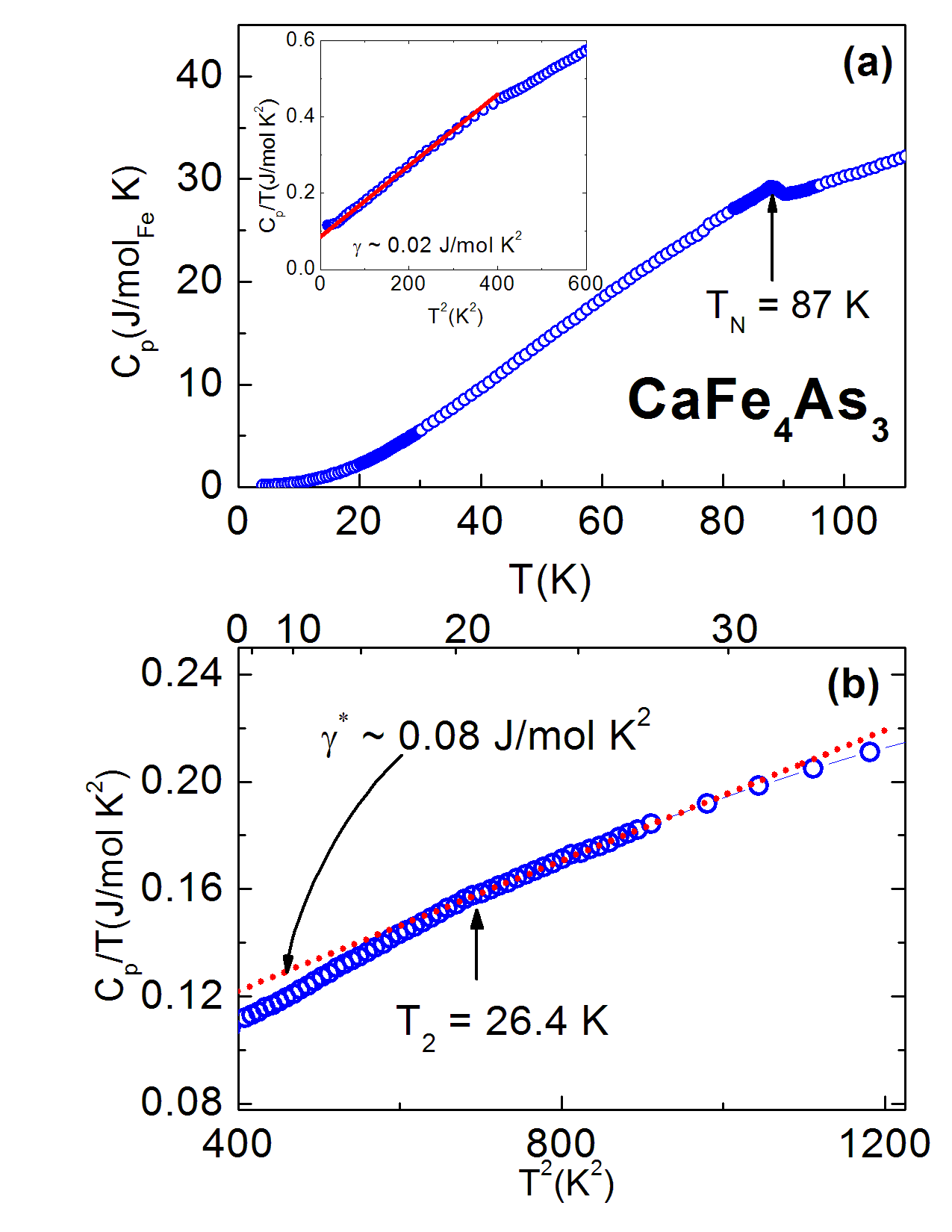}
\caption{(a) H = 0 specific heat data, with the antiferromagnetic transition at T$_N$ = 87 K marked by the vertical arrow. Inset: low temperature region, together with the expected linear fit (straight line) to C$_p/$T = $\gamma$ $+~\beta$T$^2$, which yields $\gamma$ = 0.02 J$/$(mol K$^2$). (b) C$_p/$T \textit{vs.} T$^2$ around T$_2$ = 26.4 K, and a dotted line showing a linear fit above T$_2$ with a new $\gamma^*$ = 0.08 J$/$(mol K$^2$)} \label{Cp}
\end{center}
\end{figure}

Given the resistivity behavior around T$_N$ (Fig. \ref{RT}), it is plausible that the ordering in CaFe$_4$As$_3$ may also be associated with a SDW transition: The high temperature resistivity is linearly decreasing with T, typical of a metal. Thus for simple antiferromagnetic metal, one would expect a drop in the resistivity below T$_N$, associated with loss of spin-disorder scattering. However, the resistivity of CaFe$_4$As$_3$ has a weak minimum at T$_N~\approx$ 89.0 K (Fig. \ref{RT}c), followed by a broad maximum at lower temperatures, similar to the behavior observed in Cr, a typical SDW system \cite{CrSDW}. The antiferromagnetic transition is also consistent with the H $=$ 0 specific heat data (Fig. \ref{Cp}), which shows a peak centered around 87 K. An identical C$_p$(T) measurement was recorded in an applied magnetic field H $=$ 9 T (not shown, for clarity).

A more complex phase transition is observed upon further lowering the temperature. After a decrease in $\chi$(T) below T$_N$, a sharp increase in the susceptibility occurs below 26.4 K (inset, Fig. \ref{MT}). This is associated with a similarly abrupt drop in resistivity (Fig. \ref{RT}b), followed by a much stronger decrease at the lowest measured temperatures compared to the high-T regime. However this transition cannot be detected in the specific heat data (Fig. \ref{Cp}) so a careful analysis of the low temperature properties of CaFe$_4$As$_3$ is imperative.

At least two scenarios can be considered in explaining the transition below 26.4 K: by analogy with Cr \cite{CrSDW}, a spin-flip transition is possible in the SDW state, between a transversely polarized state (where the spin direction \textbf{S} and the wavevector \textbf{q} are orthogonal) and a longitudinally polarized state (where \textbf{S} and \textbf{q} are parallel). Another possibility would be a weak structural distortion. This may be associated with the enhancement of the density of states at the Fermi surface, which in turn increases the Pauli susceptibility while correspondingly reducing the electrical resistivity. The lack of any signature of the 26.4 K transition in the specific heat data seems to render more credibility to the spin-flip hypothesis. However, as is shown below, the low temperature properties of CaFe$_4$As$_3$ are dominated by strong electron-electron coupling, which may conceal a phonon-driven feature in the specific heat at low temperatures.

Based on the susceptibility measurements, the large $\chi_0$ contributions already point to enhanced electron interactions compared to a free-electron system. Moreover, below T $<$ 20 K, the specific heat data can be fit to C$_p/$T $=$ $\gamma$ $+~\beta$T$^2$ (inset, Fig. \ref{Cp}a), which gives the  electronic and lattice specific heat coefficients $\gamma~=$ 0.02 J$/($mol$_{Fe}$K$^2$) and $\beta~=$ 0.23$*10^{-3}$ J$/($mol$_{Fe}$K$^4$) respectively. The corresponding Debye temperature is approximately $\theta_D$ = 260 K. $\gamma$ is comparable to the respective value for BaFe$_2$As$_2$ \cite{nini01}, but much larger than that reported for LaOFeAs \cite{dong01} and CaFe$_2$As$_2$ \cite{nini02} and, more importantly, significantly enhanced compared to the free electron model. An estimate of the Wilson ratio R $=~4\chi_0/(3\gamma)(\pi k_B/\mu_{eff})^2$, \cite{Wilson} gives a remarkably large value R $=$ 8 to 10 for the two applied field directions. For a non-interacting Fermi liquid, R is expected to be close to 1, and it has been shown to be closer to 2 for strongly correlated electron systems \cite{Wilson2}. Julian \emph{et al.} \cite{WilsonPd} showed that higher R values occur in nearly ferromagnetic metals. While ferromagnetic coupling in the low temperature regime may also explain the upturn in susceptibility at the lowest temperatures in CaFe$_4$As$_3$ (inset, Fig. \ref{MT}), another plausible explanation for the large Wilson ratio in this compound can be drawn from the  low temperature transition. If indeed the transition below T $<$ 26.4 K has a structural component, this would affect the electron-phonon interactions and thus $\gamma$, but would not change $\chi_0$. In this case the high Wilson ratio would be an artifact due to the fact that the $\gamma$ and $\chi_0$ values correspond to the low and high temperature states respectively. Although no phase transition can be detected in the specific heat measurement, C$_p/$T \emph{vs.} T$^2$ (Fig. \ref{Cp}b) appears to change to a linear region with lower slope and larger $\gamma$ for almost a 10 K interval above T$_2$: $\gamma^*~=$ 0.08 J$/($mol$_{Fe}$K$^2$). With both $\gamma$ and $\chi_0$ values estimated above the T$_2$ = 26.4 K transition, the Wilson ratio associated with this high temperature ordered state is R$^*~\approx~2$, a value expected for strongly correlated electron systems.

\begin{figure}[t]
\begin{center}
\includegraphics[width=2.4in]{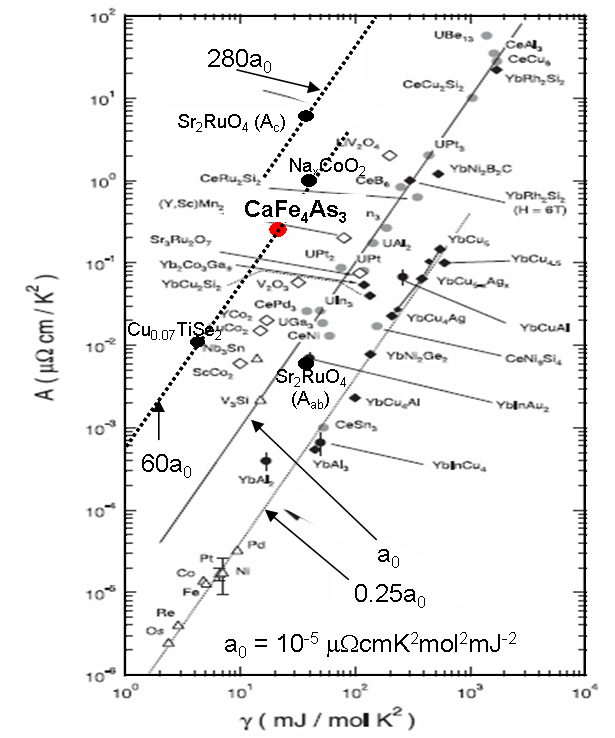}
\caption{Kadowaki-Woods plot A \textit{vs.} $\gamma$ (reproduced from \cite{KW}), with CaFe$_4$As$_3$ shown as large red symbol. CaFe$_4$As$_3$ has similar KW ratio to Na$_{0.07}$CoO$_2$ \cite{NaCobaltate} and Cu$_{0.07}$TiSe$_2$ \cite{CuTiSe2}, smaller only than that for Sr$_2$RuO$_4$ (for $\textit{i}\parallel$c) \cite{SrRuthenate}} \label{KW}
\end{center}
\end{figure}

The low temperature dependence of the resistivity (Fig. \ref{RT}b) reveals metal-like behavior with a residual resistivity $\rho_0~\approx$ 45 $\mu\Omega$ cm. By plotting $\Delta\rho$ $\emph{vs.}$ T$^2$ (Fig. \ref{RT}d), where $\Delta\rho$(T) = $\rho($T$)~-~\rho_0$ = A T$^2$, a clear Fermi-liquid regime is observed below T $\approx$ 15 K for both H $=$ 0 (crosses) and H $=$ 9 T (triangles). The coefficient A of the quadratic resistivity term is determined to be A $=$ 0.25 $\mu\Omega$ cm K$^{-2}$, remarkably large and field independent. When normalizing A by the low-temperature quasiparticle effective mass $\gamma$ = 0.02 J$/($mol$_{Fe}$K$^2$), the Kadowaki-Woods (KW) ratio A$/\gamma^2$ is almost two orders of magnitude larger than for heavy fermion materials: A$/\gamma^2$ = 55 a$_0$, where a$_0$ = 10$^{-5}$ $\mu~\Omega$cm mol$^2$ K$^2 $mJ$^{-2}$ is a nearly universal value observed in strongly correlated electron systems \cite{KW}. This value is among the largest KW ratios reported (Fig. \ref{KW}),  significantly smaller only than that of Sr$_2$RuO$_4$ for current normal to the RuO$_2$ planes \cite{SrRuthenate}. Various mechanisms have been proposed to explain the large KW ratios. Magnetic frustration or the proximity to a quantum critical point (QCP) were considered likely scenarios for the A$/\gamma^2$ = 60 a$_0$ observed in Na$_{0.07}$CoO$_2$ \cite{NaCobaltate}. In Sr$_2$RuO$_4$, the KW ratio was highly anisotropic, likely due to the two-dimensional character of the Fermi liquid, and the reduced dimensionality may be the cause for enhanced KW ratio in Cu$_{0.07}$TiSe$_2$\cite{CuTiSe2}. The fact that A and $\gamma$ are nearly field-independent up to 9 T in CaFe$_4$As$_3$ doesn't immediately justify a ``QCP" scenario, although it will be important to study the effect of pressure or doping on the possible SDW state and the low-temperature transition. The reduced dimensionality of the Fermi liquid cannot explain  the high KW ratio either, given the 3D character of the structure in CaFe$_4$As$_3$. Modest magnetic frustration may be present in this compound given that the Weiss temperatures $\theta_b$ and $\theta_{ac}$ are $\sim~-$ 220 K and $-$ 150 K respectively, 2 to 2.5 times larger than T$_N$ = 88.0 K. It will be necessary to study the microscopic magnetic structure to determine whether the Fe sublattice is indeed frustrated, and this work is currently underway.

In conclusion, CaFe$_4$As$_3$ displays a Fermi liquid behavior at low temperatures, with enhanced electron-electron correlations as indicate by the value of the electronic specific heat coefficient $\gamma~=$ 0.02 J$/($mol$_{Fe}$K$^2$), and an unusually high KW ratio A$/\gamma^2$ = 55 a$_0$. A low temperature transition exists in this compound around T$_2$ = (26.4 $\pm$ 1.0) K, marked by a remarkably sharp resistivity drop and abrupt increase in susceptibility. Experiments involving low-temperature neutron diffraction are underway to elucidate the nature of the magnetic structure below and above this transition, and to clarify whether it is magnetic in nature or it has a structural component as well. At higher temperatures, antiferromagnetic order occurs below T$_N$ = (88.0 $\pm$ 1.0) K, which is likely associated with a SDW state as suggested by the small increase in the resistivity below T$_N$. Although not superconducting and with a more 3D crystal structure than the iron pnictides, CaFe$_4$As$_3$ resembles the ``1-2-2" superconductors with regards to the high temperature SDW ordering, the enhanced electron mass $\gamma$ and also the Fe-As sheets. Together with the enhanced R and KW ratios, the room temperature resistivity values around 0.3 m$\Omega$cm, comparable to those of LaFeOP \cite{kam08jacs} and AFe$_2$As$_2$ (A = Ba \cite{nini01} or Ca \cite{nini02}), qualify the new compound CaFe$_4$As$_3$ as a metal with enhanced electron correlations.

We would like to thank Q. Si, D. Natelson and M. Dzero for useful discussions. EM acknowledges support from Rice University. Work at UC Davis was funded by NSF DMR-0600742.

\bibliographystyle{apsrev}
\bibliography{CaFe4As3_LZnew05}

\end{document}